\documentclass[sigconf]{acmart}
\AtBeginDocument{%
  \providecommand\BibTeX{{%
    \normalfont B\kern-0.5em{\scshape i\kern-0.25em b}\kern-0.8em\TeX}}}

\setcopyright{acmcopyright}
\copyrightyear{2018}
\acmYear{2018}
\acmDOI{10.1145/1122445.1122456}

\DeclareMathOperator*{\relu}{ReLU}
\setcopyright{rightsretained}
\acmConference[SIGIR eCom'21]{ACM SIGIR Workshop on eCommerce}{July 15, 2021}{Virtual Event, Montreal, Canada} \acmYear{2021} \copyrightyear{2021} \makeatletter \renewcommand\@formatdoi[1]{\ignorespaces}
\makeatother
\acmISBN{}

\begin{document}

\title{Aligning Hotel Embeddings using Domain Adaptation for Next-Item Recommendation}

\author{Ioannis Partalas}
\email{ipartalas@expediagroup.com}
\affiliation{%
  \institution{Expedia Group}
  \city{Geneva}
  \country{Switzerland}
}


\begin{abstract}
 In online platforms it is often the case to have multiple brands under the same group which may target different customer profiles, or have different domains. For example, in the hospitality domain, Expedia Group has multiple brands like Brand Expedia, Hotels.com and Wotif which have either different traveler profiles or are more relevant in a local context.
 
In this context, learning embeddings for hotels that can be leveraged in recommendation tasks in multiple brands requires to have a common embedding that can be induced using alignment approaches. In the same time, one needs to ensure that this common embedding space does not degrade the performance in any of the brands.

In this work we build upon the hotel2vec model and propose a simple regularization approach for aligning hotel embeddings of different brands via domain adaptation. We also explore alignment methods previously used in cross-lingual embeddings to align spaces of different languages. We present results on the task of next-hotel prediction using click sessions from two brands. The results show that the proposed approach can align the two embedding spaces while achieving good performance in both brands. Additionally, with respect to single-brand training we show that the proposed approach can significantly reduce training time and improve the predictive performance.
\end{abstract}




\maketitle

\section{Introduction}
Online platforms often have multiple brands, for the same line of business, under the same group which may target
different customer profiles. As an example, in the hospitality domain Expedia Group (EG) has brands like Brand Expedia, Hotels.com and Wotif, that can either have different profiles of travelers or be more relevant locally. 

A main task in such online platforms or more generally in retail platforms,
is to recommend products to customers which requires to learn an embedding space that captures their salient attributes. Hence, enable similarity comparisons that can be leveraged from recommendation systems.

In the recent years approaches that learn product embeddings from the interactions of the customers with the on-line platform have been proposed \cite{prod2vec, metaprod2vec, youtube}
as well as approached tailored to the hospitality domain \cite{hotel2vec, airbnb}. These approaches leverage the seminal word2vec model \cite{word2vec} in order to learn the embedding space by treating the clicked items as tokens in a sentence. While it is common to learn such embeddings in a single domain/brand, in the context of electronic commerce we would like to be able to leverage such embeddings across different domains/brands. As mentioned previously, one can learn hotel embeddings on Brand Expedia and leverage them in Hotels.com in order to bootstrap or improve the hotel embeddings in the latter case. Subsequently these hotel representations can be used in tasks like personalized recommendations. 

To do so, one would need to align the embedding spaces of the different brands and use this aligned space to capture the intent of the users while searching on the on-line platform. In a very recent work, Bianchi et al. \cite{coveo20} study the alignment of product embeddings to enable zero-shot learning in a cross-shop scenario. Their setting is more general as in our case the multi-brands belong to the same domain, and usually we dispose a one-to-one mapping between products (hotels in our case).

In this work we propose to align embeddings from different brands using a simple regularization approach from domain adaptation. We build upon the hotel2vec model \cite{hotel2vec} for learning hotel embeddings and extend it to accommodate alignment of  embedding spaces. Our approach, can also be thought of as a transfer learning one as we are able to boostrap the learning procedure of hotel2vec in a target brand using the hotel2vec embeddings from a source brand.

The main contributions of this paper are the following:
\begin{itemize}
    \item We propose a simple yet effective domain adaptation approach that adds a regularization term in the loss function of hotel2vec.
    \item We present empirical results on the task of next-hotel prediction for two brands.
    \item We also implement an alignment method borrowed from the task of alignment of cross-lingual embeddings. We show empirically, that to perform such a transpose to another domain one should be careful of the particularities of it.  
\end{itemize}

\section{Related Work}
In recommendation systems learning embeddings for users and items that capture the semantics is a critical part. In the domain of electronic commerce, approaches as prod2vec have been proposed in order to learn representations of products \cite{prod2vec, barkan2016item2vec, airbnb} which leverage the skip-gram model \cite{word2vec}. In \cite{youtube} the authors propose to learn embeddings for YouTube videos by combining multiple features, which are used for candidate generation and ranking. Other approaches have also been proposed that include metadata \cite{metaprod2vec} or try to capture different aspects of the product from different sources (clickstream data, text and images) \cite{singh}.

Aligning embedding spaces is a topic that has been extensively studied in the NLP domain and specifically in the context of cross-lingual embeddings. A simple and efficient approach to align two embedding spaces of different languages is a to learn a linear projection \cite{Mikolov_LP}. In \cite{artetxe-etal-2017-learning} the authors employ an iterative approach starting from seed mappings of words (source language to target language) and do the linear projection by imposing an orthogonality constraint in the projection matrix. A survey on cross-lingual word embedding can be found in \cite{ruder_survey}. In this work we study the effectiveness of such alignment approaches in the context of hotel embeddings.

Domain adaptation is also a topic that has attracted interest in the recommendation systems space \cite{darec, deep_da} where the goal is to transfer knowledge from a source to a target task. In our setting we employ a domain adaptation approach in order to align different embedding spaces and we follow a straightforward regularization approach \cite{daume-iii-2007-frustratingly}. More recently, such approaches have been encompassed in a single framework for domain adaptation \cite{lu-etal-2016-general}.
Other works, propose adversarial approaches for domain adaptation that learn in the same time the alignemnt as well as an embedding space that is invariant to the domain \cite{few_shot_da, jmlr_da, chen-etal-2018-adversarial}.

The most similar approach to our work is the one presented in \cite{coveo20} where the authors propose to align product embeddings to enable cross-shop recommendations. Specifically, they propose different approaches either based on content (images and text) or using the clicked products in the different shops as supervised signals in methods for alignment of cross-lingual embeddings. Our use case has a simpler setting as we dispose a mapping between products (hotels) in the two embeddings spaces we wish to align.
\section{Alignment through Regularized Domain Adaptation}

In order to learn representations of hotels, we use the hotel2vec model \cite{hotel2vec} which implements a neural model and is trained with the skip-gram model and negative sampling. The model learns different embeddings for click ($V_c$), hotel properties ($V_a$) and geographical information ($V_g$) which are fused to learn an enriched representation. Specifically, the embeddings are calculated as follows:

\begin{align*}
    V_c \! &= f(I_c; W_c) \\
    V_a \! &= f(I_a; W_a) \\
    V_g \! &= f(I_g; W_g)
\end{align*}

where $f(x; W) = \relu(\frac{x W}{\hphantom{{}_2}\| x W \|_{\scriptstyle{2}}})$ and $I_c, I_a$ and $I_g$ refer to the input features for the click, amenity and geographical embedding. The final hotel2vec embedding is calculated as a projection of the concatenated embeddings:
\begin{equation}
 \relu( [V_c, V_a, V_g]^\top W_e)
 \label{hotel2vec_enriched}
\end{equation}.

Let $\mathbf{V_{h_{i}}}$ be the representation for hotel $h_i$ as calculated by equation \ref{hotel2vec_enriched} where $h_i \in H$. Then hotel2vec model minimizes the following loss function:
\begin{equation}
J(\theta) = - \sum_{(h_t,h_c)\in D^+} \log\sigma\left(V_{h_t}^\intercal V_{h_c}\right) + \sum_{h_i\in N_c}\log\sigma\left(-V_{h_t}^\intercal V_{h_i}\right)
\label{eq1}
\end{equation}

where $D^+$ are the skip-gram pairs of clicked hotels in the session that are generated using a fixed length window. $N_c$ are the negative samples which are sampled from the same market of clicked hotels in the session, as a traveler searches in a specific destination. Finally, $\sigma()$ is the sigmoid function.

In this work we propose to extend the hotel2vec model in order to 
accommodate embedding spaces alignment in the case of multi-brand representations.

We do so by employing a domain adaptation scheme. Denote 
$\mathcal{D^{\textsc{s}}}$ the source domain where we already learned a hotel2vec model by minimizing the loss function in Equation \ref{eq1}, $J_s(\theta)$ . Then in the target domain
$\mathcal{D^{\textsc{t}}}$ we learn hotel2vec representations by minimizing the following regularized function:

$$
J_t(\theta) =  J(\theta) + \lambda ||V_{h_t}^{\textsc{t}} - V_{h_t}^{\textsc{s}}||_2
$$
where $V_{h_t}^{\textsc{s}}$ refers to the corresponding embedding of hotel $h_t$ in the source domain. Note that the embeddings of the source domain are fixed and not re-trained. Also, $\lambda$ is a parameter that controls how much knowledge we would like to transfer from the source domain. As in our case we want to align the embedding spaces, we would like to constraint the model to be as close as possible in the source domain, hence set $\lambda$ equal to 1.0. In the experimental section we experiment with  this parameter to understand the impact in the downstream task. 

Note that we define the strength of regularization globally but in a more fine version it could be defined per hotel. We leave this for future work.

The above regularizer assumes that we dispose a mapping between hotels in the two domains which is our case. Also, note that we do not suffer here from the problem of absent hotels from the training data in the source domain as the model can impute these ones.

\section{Experimental Setting}
We evaluate the proposed approach in the task of next-item prediction where we want to predict the next clicked hotel in the session based on the previous clicked hotel.

We collect click sessions over one year of searches for both Hotels.com and Brand Expedia. It contains more than 65M user click sessions for each brand and around 700K unique hotels for Hotels.com dataset and over 1.4M for Brand Expedia respectively. We randomly split the sessions in training, validation and test with a ration of 8:1:1.

We use a system with 64GB RAM, 8 CPU cores, and a
Tesla V100 GPU. We use Python 3 as the programming language
and the Tensorflow \cite{tensorflow2015-whitepaper} library for the
neural network architecture and gradient calculations.

For hotel2vec we follow the experimentation methodology in \cite{hotel2vec} for tuning the hyperparameters of the model. The model in both brands is trained with $L2$-regularization.

For the alignment approaches, we use as source domain Hotels.com hotel2vec embeddings and Brand Expedia as the target one. 
For the proposed approach we set the parameter $\lambda$ to 1.0 in order to force the alignment of the spaces. We also tried a value of 0.5 in order to understand the behavior of the approach.

We compare the regularization approach with the linear projection alignment approach presented in \cite{Mikolov_LP}. In that case given the embeddings that we export from the trained models in Hotels.com and Brand Expedia domains, denoted $\mathcal{V}^{\textsc{S}}$ and $\mathcal{V}^{\textsc{T}}$ respectively, we solve the following optimization problem:
$$
\min_W||\mathcal{V}^{\textsc{S}} W - \mathcal{V}^{\textsc{T}}||_{2}^2
$$

The alignment is learned only on the common hotel embeddings for the two sets of vectors in order to avoid injecting too much noise.

\paragraph{Metrics:} We compare the different approaches in terms of hits@k and MRR@k (Mean Reciprocal Rank). Hits@k measures the average number of times the correct selection (i.e. the hotels clicked by the users in a
session) appears in the top k predicted hotels.
MRR@k evaluates the average list quality of k items returned by the model.

Both metrics are calculated over the ranking induced by cosine similarity of the embedding vectors. Note that when we have a cross-brand setting this corresponds to a zero-shot learning framework.  
\section{Results}
We present in this section results for two settings: a) the zero-shot prediction of next clicked hotel where we use embeddings in a cross-brand setting and b) the supervised setting where we compare the in-brand hotel2vec model with and without domain adaptation.
\subsection{Zero-shot Prediction}
Table \ref{table:res} presents the hits@k  and MRR@k for $k\in \{10,100\}$ to predict next-clicked hotel in the test sets for both Brand Expedia and Hotels.com. 

The first two approaches, named $hotel2vec_H$ and $hotel2vec_E$, correspond to single-domain trained models with no further alignment. The results, show how the corresponding embeddings behave in a cross-brand zero-shot learning approach.

The next two approaches present the corresponding metrics for the linear projection alignment approach (LP) as well as the proposed approach that employs domain adaptation ($hotel2vec_{DA}$).

Focusing only in the single-domain approaches, we can observe that the embeddings can enable zero-shot learning in a cross-brand scenario as they can achieve good performance across all the metrics. For example, with the $hotel2vec_H$ we can achieve 0.5073 of hits@100 compared to the in-brand trained model $hotel2vec_E$ that achieves a 0.5335 hits@100. The same observation stands in the case of Hotels.com even though we notice that the gap between the different performances is larger.

Concerning the alignment approaches presented in the next two rows of the table, we notice that the LP method has a big drop in performance in both brands. As we mentioned earlier, even though we dispose a one-to-one mapping of the hotels between brands the method fails to keep the similarities in the projected space. This shows that extra care should be taken when aligning the embedding spaces, for example require the projection matrix to be orthogonal.

Regarding the proposed approach we can observe that is able to achieve good performance in both brands. Actually, it achieves the same performance with the in-domain model in Hotels.com, 0.7535 versus 0.7573 for hits@100, while there is a drop in performance for the Brand Expedia. We recall here that we set the $\lambda$ parameter that controls the transfer of knowledge to either 0.5 or 1.0 where the latter allows us to force the alignment between brands. In the case of less regularization strength ($\lambda=0.5$) we can observe that the metrics slightly improve. Smaller values of the parameter can balance between a full transfer of knowledge and in-brand learning. For the improvement in the Hotels.com brand we hypothesize that allowing for less regularization we are able to allow more in-domain knowledge to flow and learn a better similarity space. 

\begin{table*}
\caption{Hits@k and MRRk for the different approaches for embeddings alignment. Note that the hotel2vec approach refers to the standard training with no alignment for the two brands.}
\begin{tabular}{c|cccc|cccc}
\toprule
Embeddings  & \multicolumn{4}{c}{Expedia} & \multicolumn{4}{c}{Hcom} \\ \midrule
 & hits@100 & hits@10 & MRR@10 & MRR@100 & hits@100 & hits@10 & MRR@10 & MRR@100 \\\midrule
 $hotel2vec_H$ & 0.5073 &0.1593 &0.0393 &0.0519 & 0.7573 & 0.2981 & 0.1102 & 0.1276 \\
 $hotel2vec_E$ & 0.5335&0.1667 & 0.0411& 0.0545& 0.7039 & 0.2624 &0.1014 & 0.1178 \\ \midrule
 LP & 0.3593  & 0.1203 &0.0302 & 0.0389 & 0.4715 & 0.1970 & 0.0861 & 0.0962 \\
 $hotel2vec_{DA}$, $\lambda =1.0$ & 0.5015 & 0.1591 & 0.0392 & 0.0517 & 0.7535 & 0.2943 & 0.1094 & 0.1268  \\ 
 $hotel2vec_{DA}$, $\lambda =0.5$ & 0.5042 & 0.1594 & 0.0393 & 0.0519 & 0.7553 & 0.2964 & 0.1097 & 0.1271  \\ 
 
 \bottomrule
\end{tabular}
\label{table:res}
\end{table*}

\subsection{Supervised Setting}
In this section we present results when we compare the same in-brand model with and without domain adaptation. Note that in this case we let the neural model score the candidate hotels.

\begin{figure}
    \centering
    \includegraphics[width=\linewidth]{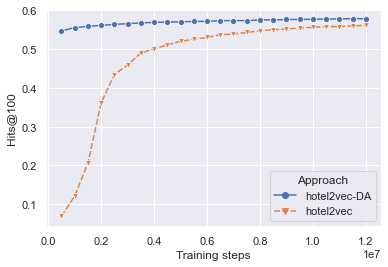}
    \caption{Hits@100 on the test set during training. We compare the domain adaptation approach with the single hotel2vec model on Brand Expedia clicks.}
    \label{fig:hits@100}
\end{figure}

\begin{figure}
    \centering
    \includegraphics[width=\linewidth]{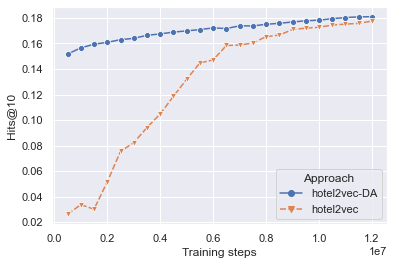}
    \caption{Hits@10 on the test test during training. We compare the domain adaptation approach with the single hotel2vec model on Brand Expedia clicks.}
    \label{fig:hits@10}
\end{figure}

Figures \ref{fig:hits@10} and \ref{fig:hits@100} present the hits@10 and hits@100 respectively for hotel2vec with domain adaptation and a single-brand model. We can notice that the proposed regularization scheme improves both the metrics as well as it can reduce training time. Also, the jump-start is significant and shows that we can share knowledge between the two brands.

Table \ref{tbl:h2vec} presents the hits@k metric for the in-brand hotel2vec model in comparison with the proposed approach when we use the model to score the next-hotel to be clicked. We can clearly see that the domain adaptation approach outperforms the single-brand approach by 3\% for hits@100 metric. 

\begin{table}
\caption{Comparison of the single-brand hotel2vec model versus the domain adaptation approach in terms of hits@k. The metrics are measured when the scores for the next-hotel prediction are produced by the model.}
\begin{tabular}{c|c|c}
\toprule
Approach     & hits@10  & hits@100 \\ \midrule
$hotel2vec_E$     & 17.76 & 56.11 \\
$hotel2vec_{DA}$ & 18.10 & 57.81 \\
\bottomrule
\end{tabular}
\label{tbl:h2vec}
\end{table}
\section{Conclusions}
We presented in this work a simple yet effective regularization approach for aligning embedding spaces in a multi-brand scenario. For example, Expedia Group has multiple brands that operate in the same domain. The idea is to add a regularizer in the objective function of the model that learns the embeddings in order to force them to be as close as possible to the embeddings of the source brand, hence performing domain adaptation. This kind of approaches has also been explored in the past in Natural Language Processing tasks
\cite{daume-iii-2007-frustratingly}.

We evaluated the proposed approach in the next-hotel prediction task for two brands, that is Expedia and Hotels.com. We measured performance in terms of hits@k and MRR@k metrics. We also, compared with linear projection alignment borrowed by the cross-lingual approaches for aligning embeddings of different languages \cite{Mikolov_LP}.

The results, showed that the proposed approach can align the spaces of the multiple brands achieving good performance in both brands. Additionally, we showed that the regularized version of the hotel2vec model can learn faster and improve the performance. We also observed that an approach like the linear projection without taking into account some particularities of the domain leads to worse performance.
\section{Future Work}
For future work we would like to add a more adaptive regularization parameter that can be defined per hotel rather than being global. In that way we may wish to transfer knowledge when we are certain that a pair of hotels in the source and target brands should have the same embedding.

We would also like to explore the use of multiple source brands in order to align in the same time the embedding spaces. Multi-task approaches can be leveraged to align the different embedding spaces \cite{lu-etal-2016-general}. Also, other alignment approaches could be explored \cite{BianchiCompass}.

Finally, we would like to explore adversarial cross-domain adaptation for aligning the embedding spaces \cite{jmlr_da}. In this case, we want to leverage the similar features across the brands while also learning specific embeddings for each brand.
\begin{acks}
We would like to thank Phong Nguyen, C\'{e}dric Houdr\'{e} and Clara Jord\'{a} Lope for discussions as well as Jan Krasnodebski and Daniele Dhongi for comments on the paper. Many thanks to Petros Baltzis for engineering support.
\end{acks}

\bibliographystyle{ACM-Reference-Format}
\bibliography{ecom_biblio}


\begin{thebibliography}{21}


\ifx \showCODEN    \undefined \def \showCODEN     #1{\unskip}     \fi
\ifx \showDOI      \undefined \def \showDOI       #1{#1}\fi
\ifx \showISBNx    \undefined \def \showISBNx     #1{\unskip}     \fi
\ifx \showISBNxiii \undefined \def \showISBNxiii  #1{\unskip}     \fi
\ifx \showISSN     \undefined \def \showISSN      #1{\unskip}     \fi
\ifx \showLCCN     \undefined \def \showLCCN      #1{\unskip}     \fi
\ifx \shownote     \undefined \def \shownote      #1{#1}          \fi
\ifx \showarticletitle \undefined \def \showarticletitle #1{#1}   \fi
\ifx \showURL      \undefined \def \showURL       {\relax}        \fi
\providecommand\bibfield[2]{#2}
\providecommand\bibinfo[2]{#2}
\providecommand\natexlab[1]{#1}
\providecommand\showeprint[2][]{arXiv:#2}

\bibitem[\protect\citeauthoryear{Abadi, Agarwal, Barham, and et. al.}{Abadi
  et~al\mbox{.}}{2015}]%
        {tensorflow2015-whitepaper}
\bibfield{author}{\bibinfo{person}{Martin Abadi}, \bibinfo{person}{Ashish
  Agarwal}, \bibinfo{person}{Paul Barham}, {and} \bibinfo{person}{et. al.}}
  \bibinfo{year}{2015}\natexlab{}.
\newblock \bibinfo{title}{{TensorFlow}: Large-Scale Machine Learning on
  Heterogeneous Systems}.
\newblock
\newblock
\urldef\tempurl%
\url{https://www.tensorflow.org/}
\showURL{%
\tempurl}
\newblock
\shownote{Software available from tensorflow.org.}


\bibitem[\protect\citeauthoryear{Artetxe, Labaka, and Agirre}{Artetxe
  et~al\mbox{.}}{2017}]%
        {artetxe-etal-2017-learning}
\bibfield{author}{\bibinfo{person}{Mikel Artetxe}, \bibinfo{person}{Gorka
  Labaka}, {and} \bibinfo{person}{Eneko Agirre}.}
  \bibinfo{year}{2017}\natexlab{}.
\newblock \showarticletitle{Learning bilingual word embeddings with (almost) no
  bilingual data}. In \bibinfo{booktitle}{\emph{Proceedings of the 55th Annual
  Meeting of the Association for Computational Linguistics (Volume 1: Long
  Papers)}}. \bibinfo{publisher}{Association for Computational Linguistics},
  \bibinfo{address}{Vancouver, Canada}, \bibinfo{pages}{451--462}.
\newblock
\urldef\tempurl%
\url{https://doi.org/10.18653/v1/P17-1042}
\showDOI{\tempurl}


\bibitem[\protect\citeauthoryear{Barkan and Koenigstein}{Barkan and
  Koenigstein}{2016}]%
        {barkan2016item2vec}
\bibfield{author}{\bibinfo{person}{Oren Barkan} {and} \bibinfo{person}{Noam
  Koenigstein}.} \bibinfo{year}{2016}\natexlab{}.
\newblock \showarticletitle{Item2vec: neural item embedding for collaborative
  filtering}. In \bibinfo{booktitle}{\emph{2016 IEEE 26th International
  Workshop on Machine Learning for Signal Processing (MLSP)}}. IEEE,
  \bibinfo{pages}{1--6}.
\newblock


\bibitem[\protect\citeauthoryear{Bianchi, Carlo, Nicoli, and Palmonari}{Bianchi
  et~al\mbox{.}}{2020a}]%
        {BianchiCompass}
\bibfield{author}{\bibinfo{person}{Federico Bianchi},
  \bibinfo{person}{Valerio~Di Carlo}, \bibinfo{person}{Paolo Nicoli}, {and}
  \bibinfo{person}{Matteo Palmonari}.} \bibinfo{year}{2020}\natexlab{a}.
\newblock \showarticletitle{Compass-aligned Distributional Embeddings for
  Studying Semantic Differences across Corpora}.
\newblock \bibinfo{journal}{\emph{CoRR}}  \bibinfo{volume}{abs/2004.06519}
  (\bibinfo{year}{2020}).
\newblock
\showeprint[arxiv]{2004.06519}
\urldef\tempurl%
\url{https://arxiv.org/abs/2004.06519}
\showURL{%
\tempurl}


\bibitem[\protect\citeauthoryear{Bianchi, Tagliabue, Yu, Bigon, and
  Greco}{Bianchi et~al\mbox{.}}{2020b}]%
        {coveo20}
\bibfield{author}{\bibinfo{person}{Federico Bianchi}, \bibinfo{person}{Jacopo
  Tagliabue}, \bibinfo{person}{Bingqing Yu}, \bibinfo{person}{Luca Bigon},
  {and} \bibinfo{person}{Ciro Greco}.} \bibinfo{year}{2020}\natexlab{b}.
\newblock \showarticletitle{Fantastic Embeddings and How to Align Them:
  Zero-Shot Inference in a Multi-Shop Scenario}.
\newblock \bibinfo{journal}{\emph{CoRR}}  \bibinfo{volume}{abs/2007.14906}
  (\bibinfo{year}{2020}).
\newblock
\showeprint[arxiv]{2007.14906}
\urldef\tempurl%
\url{https://arxiv.org/abs/2007.14906}
\showURL{%
\tempurl}


\bibitem[\protect\citeauthoryear{Chen, Sun, Athiwaratkun, Cardie, and
  Weinberger}{Chen et~al\mbox{.}}{2018}]%
        {chen-etal-2018-adversarial}
\bibfield{author}{\bibinfo{person}{Xilun Chen}, \bibinfo{person}{Yu Sun},
  \bibinfo{person}{Ben Athiwaratkun}, \bibinfo{person}{Claire Cardie}, {and}
  \bibinfo{person}{Kilian Weinberger}.} \bibinfo{year}{2018}\natexlab{}.
\newblock \showarticletitle{Adversarial Deep Averaging Networks for
  Cross-Lingual Sentiment Classification}.
\newblock \bibinfo{journal}{\emph{Transactions of the Association for
  Computational Linguistics}}  \bibinfo{volume}{6} (\bibinfo{year}{2018}),
  \bibinfo{pages}{557--570}.
\newblock
\urldef\tempurl%
\url{https://doi.org/10.1162/tacl_a_00039}
\showDOI{\tempurl}


\bibitem[\protect\citeauthoryear{Covington, Adams, and Sargin}{Covington
  et~al\mbox{.}}{2016}]%
        {youtube}
\bibfield{author}{\bibinfo{person}{Paul Covington}, \bibinfo{person}{Jay
  Adams}, {and} \bibinfo{person}{Emre Sargin}.}
  \bibinfo{year}{2016}\natexlab{}.
\newblock \showarticletitle{Deep Neural Networks for YouTube Recommendations}.
  In \bibinfo{booktitle}{\emph{Proceedings of the 10th ACM Conference on
  Recommender Systems}} (Boston, Massachusetts, USA)
  \emph{(\bibinfo{series}{RecSys '16})}. \bibinfo{publisher}{ACM},
  \bibinfo{address}{New York, NY, USA}, \bibinfo{pages}{191--198}.
\newblock
\showISBNx{978-1-4503-4035-9}
\urldef\tempurl%
\url{https://doi.org/10.1145/2959100.2959190}
\showDOI{\tempurl}


\bibitem[\protect\citeauthoryear{Daum{\'e}~III}{Daum{\'e}~III}{2007}]%
        {daume-iii-2007-frustratingly}
\bibfield{author}{\bibinfo{person}{Hal Daum{\'e}~III}.}
  \bibinfo{year}{2007}\natexlab{}.
\newblock \showarticletitle{Frustratingly Easy Domain Adaptation}. In
  \bibinfo{booktitle}{\emph{Proceedings of the 45th Annual Meeting of the
  Association of Computational Linguistics}}. \bibinfo{publisher}{Association
  for Computational Linguistics}, \bibinfo{address}{Prague, Czech Republic},
  \bibinfo{pages}{256--263}.
\newblock
\urldef\tempurl%
\url{https://www.aclweb.org/anthology/P07-1033}
\showURL{%
\tempurl}


\bibitem[\protect\citeauthoryear{Ganin, Ustinova, Ajakan, Germain, Larochelle,
  Laviolette, Marchand, and Lempitsky}{Ganin et~al\mbox{.}}{2016}]%
        {jmlr_da}
\bibfield{author}{\bibinfo{person}{Yaroslav Ganin}, \bibinfo{person}{Evgeniya
  Ustinova}, \bibinfo{person}{Hana Ajakan}, \bibinfo{person}{Pascal Germain},
  \bibinfo{person}{Hugo Larochelle}, \bibinfo{person}{Fran\c{c}ois Laviolette},
  \bibinfo{person}{Mario Marchand}, {and} \bibinfo{person}{Victor Lempitsky}.}
  \bibinfo{year}{2016}\natexlab{}.
\newblock \showarticletitle{Domain-Adversarial Training of Neural Networks}.
\newblock \bibinfo{journal}{\emph{J. Mach. Learn. Res.}} \bibinfo{volume}{17},
  \bibinfo{number}{1} (\bibinfo{date}{Jan.} \bibinfo{year}{2016}),
  \bibinfo{pages}{2096–2030}.
\newblock
\showISSN{1532-4435}


\bibitem[\protect\citeauthoryear{Grbovic}{Grbovic}{2018}]%
        {airbnb}
\bibfield{author}{\bibinfo{person}{Cheng Grbovic}.}
  \bibinfo{year}{2018}\natexlab{}.
\newblock \showarticletitle{Real-time Personalization using Embeddings for
  Search Ranking at Airbnb}. In \bibinfo{booktitle}{\emph{Proceedings of the
  24th ACM SIGKDD International Conference on Knowledge Discovery and Data
  Mining}}. KDD, \bibinfo{pages}{311--320}.
\newblock


\bibitem[\protect\citeauthoryear{Grbovic, Radosavljevic, Djuric, Bhamidipati,
  Savla, Bhagwan, and Sharp}{Grbovic et~al\mbox{.}}{2015}]%
        {prod2vec}
\bibfield{author}{\bibinfo{person}{Mihajlo Grbovic}, \bibinfo{person}{Vladan
  Radosavljevic}, \bibinfo{person}{Nemanja Djuric}, \bibinfo{person}{Narayan
  Bhamidipati}, \bibinfo{person}{Jaikit Savla}, \bibinfo{person}{Varun
  Bhagwan}, {and} \bibinfo{person}{Doug Sharp}.}
  \bibinfo{year}{2015}\natexlab{}.
\newblock \showarticletitle{E-commerce in your inbox: Product recommendations
  at scale}. In \bibinfo{booktitle}{\emph{Proceedings of the 21th ACM SIGKDD
  international conference on knowledge discovery and data mining}}.
  \bibinfo{pages}{1809--1818}.
\newblock


\bibitem[\protect\citeauthoryear{Kanagawa, Kobayashi, Shimizu, Tagami, and
  Suzuki}{Kanagawa et~al\mbox{.}}{2019}]%
        {deep_da}
\bibfield{author}{\bibinfo{person}{Heishiro Kanagawa}, \bibinfo{person}{Hayato
  Kobayashi}, \bibinfo{person}{Nobuyuki Shimizu}, \bibinfo{person}{Yukihiro
  Tagami}, {and} \bibinfo{person}{Taiji Suzuki}.}
  \bibinfo{year}{2019}\natexlab{}.
\newblock \showarticletitle{Cross-Domain Recommendation via Deep Domain
  Adaptation}. In \bibinfo{booktitle}{\emph{Advances in Information
  Retrieval}}, \bibfield{editor}{\bibinfo{person}{Leif Azzopardi},
  \bibinfo{person}{Benno Stein}, \bibinfo{person}{Norbert Fuhr},
  \bibinfo{person}{Philipp Mayr}, \bibinfo{person}{Claudia Hauff}, {and}
  \bibinfo{person}{Djoerd Hiemstra}} (Eds.). \bibinfo{publisher}{Springer
  International Publishing}, \bibinfo{address}{Cham}, \bibinfo{pages}{20--29}.
\newblock


\bibitem[\protect\citeauthoryear{Lu, Chieu, and L{\"o}fgren}{Lu
  et~al\mbox{.}}{2016}]%
        {lu-etal-2016-general}
\bibfield{author}{\bibinfo{person}{Wei Lu}, \bibinfo{person}{Hai~Leong Chieu},
  {and} \bibinfo{person}{Jonathan L{\"o}fgren}.}
  \bibinfo{year}{2016}\natexlab{}.
\newblock \showarticletitle{A General Regularization Framework for Domain
  Adaptation}. In \bibinfo{booktitle}{\emph{Proceedings of the 2016 Conference
  on Empirical Methods in Natural Language Processing}}.
  \bibinfo{publisher}{Association for Computational Linguistics},
  \bibinfo{address}{Austin, Texas}, \bibinfo{pages}{950--954}.
\newblock
\urldef\tempurl%
\url{https://doi.org/10.18653/v1/D16-1095}
\showDOI{\tempurl}


\bibitem[\protect\citeauthoryear{Mikolov, Le, and Sutskever}{Mikolov
  et~al\mbox{.}}{2013a}]%
        {Mikolov_LP}
\bibfield{author}{\bibinfo{person}{Tom{\'{a}}s Mikolov},
  \bibinfo{person}{Quoc~V. Le}, {and} \bibinfo{person}{Ilya Sutskever}.}
  \bibinfo{year}{2013}\natexlab{a}.
\newblock \showarticletitle{Exploiting Similarities among Languages for Machine
  Translation}.
\newblock \bibinfo{journal}{\emph{CoRR}}  \bibinfo{volume}{abs/1309.4168}
  (\bibinfo{year}{2013}).
\newblock
\showeprint[arxiv]{1309.4168}
\urldef\tempurl%
\url{http://arxiv.org/abs/1309.4168}
\showURL{%
\tempurl}


\bibitem[\protect\citeauthoryear{Mikolov, Sutskever, Chen, Corrado, and
  Dean}{Mikolov et~al\mbox{.}}{2013b}]%
        {word2vec}
\bibfield{author}{\bibinfo{person}{Tomas Mikolov}, \bibinfo{person}{Ilya
  Sutskever}, \bibinfo{person}{Kai Chen}, \bibinfo{person}{Greg~S Corrado},
  {and} \bibinfo{person}{Jeff Dean}.} \bibinfo{year}{2013}\natexlab{b}.
\newblock \showarticletitle{Distributed representations of words and phrases
  and their compositionality}. In \bibinfo{booktitle}{\emph{Advances in neural
  information processing systems}}. \bibinfo{pages}{3111--3119}.
\newblock


\bibitem[\protect\citeauthoryear{Motiian, Jones, Iranmanesh, and
  Doretto}{Motiian et~al\mbox{.}}{2017}]%
        {few_shot_da}
\bibfield{author}{\bibinfo{person}{Saeid Motiian}, \bibinfo{person}{Quinn
  Jones}, \bibinfo{person}{Seyed Iranmanesh}, {and} \bibinfo{person}{Gianfranco
  Doretto}.} \bibinfo{year}{2017}\natexlab{}.
\newblock \showarticletitle{Few-Shot Adversarial Domain Adaptation}. In
  \bibinfo{booktitle}{\emph{Advances in Neural Information Processing
  Systems}}, \bibfield{editor}{\bibinfo{person}{I.~Guyon},
  \bibinfo{person}{U.~V. Luxburg}, \bibinfo{person}{S.~Bengio},
  \bibinfo{person}{H.~Wallach}, \bibinfo{person}{R.~Fergus},
  \bibinfo{person}{S.~Vishwanathan}, {and} \bibinfo{person}{R.~Garnett}}
  (Eds.), Vol.~\bibinfo{volume}{30}. \bibinfo{publisher}{Curran Associates,
  Inc.}
\newblock
\urldef\tempurl%
\url{https://proceedings.neurips.cc/paper/2017/file/21c5bba1dd6aed9ab48c2b34c1a0adde-Paper.pdf}
\showURL{%
\tempurl}


\bibitem[\protect\citeauthoryear{Ruder}{Ruder}{2017}]%
        {ruder_survey}
\bibfield{author}{\bibinfo{person}{Sebastian Ruder}.}
  \bibinfo{year}{2017}\natexlab{}.
\newblock \showarticletitle{A survey of cross-lingual embedding models}.
\newblock \bibinfo{journal}{\emph{CoRR}}  \bibinfo{volume}{abs/1706.04902}
  (\bibinfo{year}{2017}).
\newblock
\showeprint[arxiv]{1706.04902}
\urldef\tempurl%
\url{http://arxiv.org/abs/1706.04902}
\showURL{%
\tempurl}


\bibitem[\protect\citeauthoryear{Sadeghian, Minaee, Partalas, Li, Wang, and
  Cowan}{Sadeghian et~al\mbox{.}}{2019}]%
        {hotel2vec}
\bibfield{author}{\bibinfo{person}{Ali Sadeghian}, \bibinfo{person}{Shervin
  Minaee}, \bibinfo{person}{Ioannis Partalas}, \bibinfo{person}{Xinxin Li},
  \bibinfo{person}{Daisy~Zhe Wang}, {and} \bibinfo{person}{Brooke Cowan}.}
  \bibinfo{year}{2019}\natexlab{}.
\newblock \showarticletitle{Hotel2vec: Learning Attribute-Aware Hotel
  Embeddings with Self-Supervision}.
\newblock \bibinfo{journal}{\emph{CoRR}}  \bibinfo{volume}{abs/1910.03943}
  (\bibinfo{year}{2019}).
\newblock
\showeprint[arxiv]{1910.03943}
\urldef\tempurl%
\url{http://arxiv.org/abs/1910.03943}
\showURL{%
\tempurl}


\bibitem[\protect\citeauthoryear{Singh, Singh, Arora, and Borar}{Singh
  et~al\mbox{.}}{2019}]%
        {singh}
\bibfield{author}{\bibinfo{person}{Loveperteek Singh}, \bibinfo{person}{Shreya
  Singh}, \bibinfo{person}{Sagar Arora}, {and} \bibinfo{person}{Sumit Borar}.}
  \bibinfo{year}{2019}\natexlab{}.
\newblock \showarticletitle{One Embedding To Do Them All}.
\newblock \bibinfo{journal}{\emph{arXiv preprint arXiv:1906.12120}}
  (\bibinfo{year}{2019}).
\newblock


\bibitem[\protect\citeauthoryear{Vasile, Smirnova, and Conneau}{Vasile
  et~al\mbox{.}}{2016}]%
        {metaprod2vec}
\bibfield{author}{\bibinfo{person}{Flavian Vasile}, \bibinfo{person}{Elena
  Smirnova}, {and} \bibinfo{person}{Alexis Conneau}.}
  \bibinfo{year}{2016}\natexlab{}.
\newblock \showarticletitle{Meta-prod2vec: Product embeddings using
  side-information for recommendation}. In
  \bibinfo{booktitle}{\emph{Proceedings of the 10th ACM Conference on
  Recommender Systems}}. ACM, \bibinfo{pages}{225--232}.
\newblock


\bibitem[\protect\citeauthoryear{Yuan, Yao, and Benatallah}{Yuan
  et~al\mbox{.}}{2019}]%
        {darec}
\bibfield{author}{\bibinfo{person}{Feng Yuan}, \bibinfo{person}{Lina Yao},
  {and} \bibinfo{person}{Boualem Benatallah}.} \bibinfo{year}{2019}\natexlab{}.
\newblock \showarticletitle{DARec: Deep Domain Adaptation for Cross-Domain
  Recommendation via Transferring Rating Patterns}. In
  \bibinfo{booktitle}{\emph{Proceedings of the Twenty-Eighth International
  Joint Conference on Artificial Intelligence, {IJCAI-19}}}.
  \bibinfo{publisher}{International Joint Conferences on Artificial
  Intelligence Organization}, \bibinfo{pages}{4227--4233}.
\newblock
\urldef\tempurl%
\url{https://doi.org/10.24963/ijcai.2019/587}
\showDOI{\tempurl}


\end{thebibliography}


\end{document}